\documentclass[12pt]{article}
\usepackage{amssymb}
\usepackage{myart}
\usepackage[dvipdf]{graphicx}

\normalbaselines
\input tcilatex
\begin{document}

\title{Non-Riemannian model of the space-time responsible for quantum effects%
}
\author{Yuri A.Rylov}
\date{Institute for Problems in Mechanics, Russian Academy of Sciences,\\
101-1, Vernadskii Ave., Moscow, 119526, Russia.\\
e-mail: rylov@ipmnet.ru\\
Web site: {$http://rsfq1.physics.sunysb.edu/\symbol{126}rylov/yrylov.htm$}\\
or mirror Web site: {$http://gasdyn-ipm.ipmnet.ru/\symbol{126}%
rylov/yrylov.htm$}}
\maketitle

\begin{abstract}
A class of homogeneous isotropic space-time models including
pseudo-Euclidean space as a special case is considered. Such a model is
chosen, where the particle motion is described in the most adequate way. It
means that the world tubes of all free particles (both classical and
quantal) are characteristic geometrical structures of the space-time. The
world function $\sigma $ of this model has the form $\sigma =\sigma _{%
\mathrm{E}}+\sigma _{0}$, $\sigma _{\mathrm{E}}>\sigma _{0}$ , where $\sigma
_{\mathrm{E}}$ is the world function of the pseudo-Euclidean space, $\sigma
_{0}$ is a constant responsible for quantum effects. It is proportional to
Planck's constant $\hbar $.
\end{abstract}

\section{Introduction}

When the gravitation is neglected, then the four-dimensional
pseudo-Euclidean space is used as a model of the space-time. It can be
motivated as follows. First, some of characteristic geometric structures of
the pseudo-Euclidean space (straight lines) occur to be real physical
objects. For instance, world lines of free macroscopic particles are
timelike straights, i.e., characteristic geometrical structures of the
pseudo-Euclidean space. Second, we have no homogeneous models of the
space-time other than pseudo-Euclidean space and the constant curvature
space.

But world lines of free microparticles (electrons, positrons, etc. ) are
stochastic. They are not characteristic structures of the pseudo-Euclidean
space. A description of the microparticle motion needs use of the quantum
mechanics principles. Is it possible such an isotropic homogeneous model of
the space-time, that its characteristic geometric structures (analog of
timelike straights) would describe the motion of both macro- and
microparticles? If so, then the quantum features of the particle motion
would be found in the spacetime model in itself, the quantum constant $\hbar
$ being one of the parameters of this model. Such a model would be
attractive, as far as it would not need a use of quantum principles.
Construction of such a model needs rather wide class of isotropic
homogeneous space-time models, what, in turn, needs to be beyond of scope of
the Riemannian geometry. Recently \cite{R90} such a rather strong
generalization of the Riemannian geometry appeared. This generalization
permits us to construct a rather wide class of isotropic homogeneous models
and to realize the above program. It is the $\sigma $ spaces defined as
follows.

Definition 1.1: The $\sigma $ space $V=(\Omega ,\sigma )$ is a set $\Omega $
of points $P$ with a real function $\sigma $ of any pair of points $P,Q\in
\Omega $.

The function $\sigma $ has the properties

\begin{equation}
\sigma \left( P,P\right) =0,\qquad \sigma \left( P,Q\right) =\sigma \left(
Q,P\right) ,\qquad P,Q\in \Omega  \label{a1.1}
\end{equation}%
It is called the world function or merely $\sigma $ function.

Interval%
\begin{equation}
S\left( P,Q\right) =\sqrt{2\sigma \left( P,Q\right) }=\left\{
\begin{array}{c}
\sqrt{2\sigma \left( P,Q\right) },\quad \sigma \left( P,Q\right) \geq 0 \\
i\sqrt{2\sigma \left( P,Q\right) },\quad \sigma \left( P,Q\right) <0%
\end{array}%
\right.  \label{a1.2}
\end{equation}%
between points $P,Q$ is called timelike, spacelike or null, if
correspondently the following conditions $\sigma (P,Q)>0$, $\sigma (P,Q)<0$,
or $\sigma (P,Q)=0\wedge P\neq Q$ are fulfilled.

As a rule a set $\mathcal{P}^{n}=P\left\{ ,,P,,...,P\right\} \subset \Omega $
of different points $P_{0},P_{1},...,P_{n}$ , determines a geometrical
object $\mathcal{T}(\mathcal{P}^{n})$ of the space $V$. This object is
called the tube of $n$th order. It is an analog of $n$-dimensional plane
determined by $n+1$ different points in the Euclidean space.

Definition of the $\sigma $ space realizes the simple physical idea that the
space-time properties are described completely by giving interval for every
pair of points. It agrees entirely with the conventional definition of
geometry (for instance, Riemannian geometry). But at the conventional
approach the relations of type (\ref{a1.1}) are given on a manifold $\mathbb{%
M}$ of a definite dimension (not on an arbitrary set $\Omega $). On the
manifold $\mathbb{M}$ such concepts as continuity, curve, surface, and
coordinate system are defined. In other words, at the conventional approach
both affine relations realized in the concept of manifold and metric
relations of type ( \ref{a1.1}) are introduced together. They have to be
agreed. Only in this case the Riemannian geometry arises.

First, the world function was introduced by Ruse \cite{R31a,R31b} and Synge
\cite{S31} for describing the Riemannian space. Now it is used mainly in
quantum gravitation \cite{C76,BV85}. In all cases it was used as some
derivative structure, i.e., a manifold $\mathbb{M}$ and a coordinate system $%
K$ on $\mathbb{M}$ were defined first. Then a metric tensor is determined in
this coordinate system. Thereafter, the world function is defined as a half
of square of interval measured along geodesic.

In the $\sigma $ space the affine properties introduced by means of a
manifold happen to be derivatives of metric properties \cite{R90} described
by the world function. In particular, if $\sigma $ space satisfies some
conditions (formulated in terms of $\sigma $ function), then the $\sigma $
space is a Riemannian space. It means that the set $\Omega $ is a manifold,
and one can determine its dimension in terms of $\sigma $ function,
construct a coordinate system, geodesics, etc. For the Riemannian space the
tubes $\mathcal{T}\left( \mathcal{P}^{n}\right) $ of $n$th order are $n$%
-dimensional geodesic surfaces.

If the $\sigma $ space does not satisfy these conditions, it is not a
Riemannian space. In this case some generalization of the Riemannian
geometry arises. It can be used as a model of the space-time.

Presentation of the $\sigma $ space properties and the proof of the $\sigma $
function self-sufficiency can be found in \cite{R73}. Here, we shall
demonstrate only how to construct a timelike tube $\mathcal{T}_{P_{0}P_{1}}$
of the first order in an arbitrary $\sigma $ space. Such a tube is an analog
of the timelike straight line in the pseudo-Euclidean space. Further, the
timelike tube $\mathcal{T}_{P_{0}P_{1}}$, $\sigma \left( P_{0},P_{1}\right) $
will be considered as a world tube of a real free particle, i.e., as a
geometrical object describing the real particle behavior.

\textit{Definition 1.2}: Vector $\mathbf{P}_{0}\mathbf{P}$ is an ordered set
$\{P_{0},P\}$ of two points $P_{0},P\in \Omega $. $P_{0}\in \Omega $ is the
origin and $P\in \Omega $ is the end of the vector.

\textit{Definition 1. 3}: Scalar product $(\mathbf{P}_{0}\mathbf{P\cdot P}%
_{0}\mathbf{Q})$ of two vectors $\left( {}\right) \mathbf{P}_{0}\mathbf{P}$%
\textbf{, }$\mathbf{P}_{0}\mathbf{Q}$ is the number%
\begin{equation}
\left( \mathbf{P}_{0}\mathbf{P\cdot P}_{0}\mathbf{Q}\right) =\sigma \left(
P_{0},P\right) +\sigma \left( P_{0},Q\right) -\sigma \left( P,Q\right)
\label{a1.3}
\end{equation}%
If $P=Q$, then the quantity $\left\vert \mathbf{P}_{0}\mathbf{P}\right\vert $
defined by the relation%
\begin{equation}
\left\vert \mathbf{P}_{0}\mathbf{P}\right\vert ^{2}=\left( \mathbf{P}_{0}%
\mathbf{P\cdot P}_{0}\mathbf{P}\right) =2\sigma \left( P_{0},P\right)
\label{a1.4}
\end{equation}%
is referred to as the length of the vector $\mathbf{P}_{0}\mathbf{P}$. The
length of a timelike vector ($\sigma \left( P_{0},P\right) >0$) is positive
and that of a spacelike vector ($\sigma \left( P_{0},P\right) <0$) is
imaginary.

\textit{Definition 1.4}: The vector $\mathbf{P}_{0}\mathbf{P}$ is parallel
to timelike vector $\mathbf{P}_{0}\mathbf{P}_{1}$, $(\mathbf{P}_{0}\mathbf{P}%
\upuparrows \mathbf{P}_{0}\mathbf{P)}$, if%
\begin{equation}
\left( \mathbf{P}_{0}\mathbf{P\cdot P}_{0}\mathbf{P}_{1}\right) =\left\vert
\mathbf{P}_{0}\mathbf{P}\right\vert \cdot \left\vert \mathbf{P}_{0}\mathbf{P}%
_{1}\right\vert  \label{a1.5}
\end{equation}

\textit{Definition 1.5}: Vector $\mathbf{P}_{0}\mathbf{P}$ is antiparallel
to the timelike vector $\mathbf{P}_{0}\mathbf{P}_{1}$ $\left( \mathbf{P}_{0}%
\mathbf{P\uparrow \downarrow P}_{0}\mathbf{P}_{1}\right) $, if%
\begin{equation}
\left( \mathbf{P}_{0}\mathbf{P\cdot P}_{0}\mathbf{P}_{1}\right) =-\left\vert
\mathbf{P}_{0}\mathbf{P}\right\vert \cdot \left\vert \mathbf{P}_{0}\mathbf{P}%
_{1}\right\vert  \label{a1.6}
\end{equation}

\textit{Definition 1.6}: Timelike vectors $\mathbf{P}_{0}\mathbf{P}_{1}$,
and $\mathbf{P}_{0}\mathbf{P}$ are collinear $\left( \mathbf{P}_{0}\mathbf{P}%
\parallel \mathbf{P}_{0}\mathbf{P}_{1}\right) $, if they are either parallel
or antiparallel, that means%
\begin{equation}
F_{2}\left( P_{0},P_{1},P_{2}\right) \equiv \left\vert
\begin{array}{cc}
\left( \mathbf{P}_{0}\mathbf{P}_{1}\mathbf{\cdot P}_{0}\mathbf{P}_{1}\right)
& \left( \mathbf{P}_{0}\mathbf{P}_{1}\mathbf{\cdot P}_{0}\mathbf{P}\right)
\\
\left( \mathbf{P}_{0}\mathbf{P\cdot P}_{0}\mathbf{P}_{1}\right) & \left(
\mathbf{P}_{0}\mathbf{P\cdot P}_{0}\mathbf{P}\right)%
\end{array}%
\right\vert  \label{a1.7}
\end{equation}%
The timelike vector $\mathbf{P}_{0}\mathbf{P}_{1}$ determines the following
sets of points $P$%
\begin{equation}
\mathcal{T}_{P_{0}P_{1}}=\left\{ P|\mathbf{P}_{0}\mathbf{P}\parallel \mathbf{%
P}_{0}\mathbf{P}_{1}\right\}  \label{a1.8}
\end{equation}%
\begin{equation}
\mathcal{T}_{[P_{0}P_{1}}=\left\{ P|\mathbf{P}_{0}\mathbf{P}\upuparrows
\mathbf{P}_{0}\mathbf{P}_{1}\right\} ,\qquad \mathcal{T}_{P_{0}]P_{1}}=\left%
\{ P|\mathbf{P}_{0}\mathbf{P}\uparrow \downarrow \mathbf{P}_{0}\mathbf{P}%
_{1}\right\}  \label{a1.9}
\end{equation}%
\begin{equation}
\mathcal{T}_{[P_{0}P_{1}]}=\left\{ P|S\left( P_{0},P\right) +S\left(
P,P_{1}\right) =S\left( P_{0},P_{1}\right) \right\}  \label{a1.10}
\end{equation}%
Here $\mathcal{T}_{P_{0}P_{1}}$ is the tube of the first order. $\mathcal{T}%
_{[P_{0}P_{1}}$ and $\mathcal{T}_{P_{0}]P_{1}}$ are two tube rays, and $%
\mathcal{T}_{[P_{1}P_{2}]}$ is the tube segment consisting of points that
lie on $\mathcal{T}_{P_{0}P_{1}}$, between the points $P_{0},P_{1}$.

\section{The tubular model of the space-time}

Now let us construct the timelike tube in some special $\sigma $ space $V_{4}
$, which is defined as follows. Let $E_{4}=\left( \mathbb{M}_{4},\sigma _{%
\mathrm{E}}\right) $ be four-dimensional pseudo-Euclidean space of index 1.
Hence, the set $\mathbb{M}_{4}$ is a manifold, and $\sigma _{\mathrm{E}}$ is
the world function of the pseudo-Euclidean space. In the Galilean coordinate
system it can be presented in the form%
\begin{equation}
\sigma _{\mathrm{E}}\left( P,P^{\prime }\right) =\sigma _{\mathrm{E}}\left(
x,x^{\prime }\right) =\frac{1}{2}g_{ik}\left( x^{i}-x^{\prime i}\right)
\left( x^{k}-x^{\prime k}\right) ,\qquad S_{\mathrm{E}}=\sqrt{2\sigma _{%
\mathrm{E}}}  \label{a2.1}
\end{equation}%
\begin{equation}
g_{ik}=\text{diag}\left( c^{2},-1-1,-1\right)   \label{a2.2}
\end{equation}%
Now let us define $\sigma $ space $V_{4}=\left( \mathbb{M}_{4},\sigma _{%
\mathrm{E}}\right) $ on the set $\mathbb{M}_{4}$ with the world function%
\begin{equation}
\sigma =f\left( \sigma _{\mathrm{E}}\right)   \label{a2.3}
\end{equation}%
where $f$ is some real function (mapping function). We shall use another
function $g$ which is connected with the function $f$. It is defined by the
formula%
\[
S_{\mathrm{E}}=g\left( S\right) =\left( 1-\alpha \left( S\right) \right)
S,\qquad S=\sqrt{2\sigma },\qquad S_{\mathrm{E}}=\sqrt{2\sigma _{\mathrm{E}}}
\]%
\begin{equation}
\lim_{S\rightarrow +\infty }\frac{g^{2}\left( S\right) }{S^{2}}=1,\qquad
\lim_{S\rightarrow 0}\frac{g^{2}\left( S\right) }{S^{2}}<\infty
\label{a2.4}
\end{equation}%
where $\alpha $ is another function which describes a deflection of the $%
\sigma $ space $V_{4}$ from the pseudo-Euclidean space $E_{4}$. The $\sigma $
space $V_{4}$, is a non-Riemannian space that will be referred to as a
conformally pseudo-Euclidean $\sigma $ space. $V_{4}$ , will be treated as a
model (tubular model) of the real space-time. If $\alpha $ = 0, then $V_{4}$%
, coincides with $E_{4}$. In this case the timelike tubes degenerate into
straight lines, and the tubular model of the space-time is converted into
the pseudo-Euclidean space that will be referred to as a linear model. The
pseudo-Euclidean space $E_{4}$ is a space associated with the $\sigma $
space $V_{4}$,. $E_{4}$ will not be treated as real space-time. $E_{4}$ is
an \textit{accessory space} that is used for establishing correspondence
between description in $V_{4}$, and the conventional description in the
pseudo-Euclidean space-time of the special relativity. For instance, if $%
V_{4}$ is a real space-time, then timelike straight lines in $E_{4}$ cannot
be treated as world lines of real particles. They are only auxiliary
constructions that are useful for treating the real world tubes of the
tubular model $V_{4}$, from the standpoint of the linear one.

Let $\mathbf{P}_{0}\mathbf{P}_{1}$, be timelike vector on $\mathbb{M}_{4}$,
and $\mu =\sqrt{2\sigma \left( P_{0},P_{1}\right) }$ is its length in $V_{4}$%
, . Then its length $\mu _{\mathrm{E}}$ in $E_{4}$ is%
\begin{equation}
\mu _{\mathrm{E}}=\sqrt{2\sigma _{\mathrm{E}}\left( P_{0}P_{1}\right) }%
=g\left( \mu \right) =\mu \left( 1-\alpha \left( \mu \right) \right)
\label{a2.5}
\end{equation}%
Using Eqs. (\ref{a1.6}), (\ref{a1.2}), (\ref{a2.4}), (\ref{a2.1}), one can
obtain the equation determining the shape of the tube $\mathcal{T}%
_{P_{0}P_{1}}$ in $V_{4}$.
\newpage

\begin{figure}[ptb]
\begin{center}
\includegraphics [height=7cm,keepaspectratio]{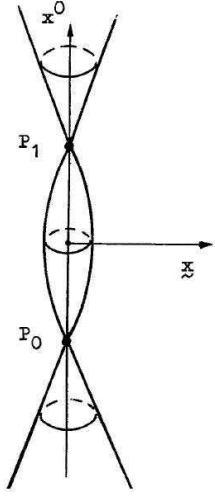}
\caption{ Schematic shape of timelike tube }\label{f1}
\end{center}
\end{figure}

Let $P_{0}=\left( -\mu _{\mathrm{E}}/2,\mathbf{0}\right) $, $P,=(\mu _{%
\mathrm{E}}/2,\mathbf{0})$ be coordinates of the points $P_{0},P_{1}$ in the
Galilean coordinate system (\ref{a2.1})) (\ref{a2.2}). Then the timelike
tube $\mathcal{T}_{P_{0}P_{1}}$ in $V_{4}$, has the shape, presented
schematically in Fig. 1. It is a three-dimensional rotation surface that
tends asymptotically to the cone surface:%
\begin{equation}
\mathbf{x}^{2}=\left( x^{0}\right) ^{2}\tanh ^{2}\theta ,\qquad \tanh \theta
=\sqrt{\left( \mu ^{2}-g^{2}\left( \mu \right) \right) /g^{2}\left( \mu
\right) }  \label{a2.6}
\end{equation}%
Section of the tube by the plane $x^{0}=0$ gives a two-dimensional sphere in
$E_{4}$ of the radius%
\begin{equation}
R_{\mathrm{E}}=\sqrt{\frac{1}{4}g^{2}\left( \mu \right) -g^{2}\left( \mu
/2\right) }  \label{a2.7}
\end{equation}%
Thus the shape of the tube $\mathcal{T}_{P_{0}P_{1}}$ depends essentially on
the length $\mu $ of the vector $\mathbf{P}_{0}\mathbf{P}_{1}$,. If $\mu $
is large enough, and hence $\mu ^{2}-g^{2}\left( \mu \right) $, then the
angle $\theta $ determining aperture of the cone, and the ratio $R_{\mathrm{E%
}}/\mu $ are small, and the tube $\mathcal{T}_{P_{0}P_{1}}$distinguishes
slightly from the straight $\mathbf{x}=0$, describing the tube $\mathcal{T}%
_{P_{0}P_{1}}$ in $E_{4}$. The less the length $\mu $ of the vector $\mathbf{%
P}_{0}\mathbf{P}_{1}$, the more aperture angle of the cone. At $\mu
\rightarrow 0$ the asymptotic cone tends to the light cone, because
according to Eq. (\ref{a2.4}) $\tanh \theta =1$ at $\mu =0$.

In classical mechanics, the four-momentum $p_{i}$,$(p_{i}p^{i}=m^{2}c^{2})$
given at the time moment $t$ \textit{determines the world line} of the free
particle. It is a straight line tangent to four-vector $p_{i}$, the world
line being independent on the particle mass. Let us treat the timelike
vector $\mathbf{P}_{0}\mathbf{P}_{1}$, as a momentum vector and its length
in the tubular model $V_{4}$, as a mass of the particle, the tube $\mathcal{T%
}_{P_{0}P_{1}}$ in $V_{4}$, being treated as the world tube of free
particle. Such a treatment is possible, because the timelike vector $\mathbf{%
P}_{0}\mathbf{P}_{1}$ determines the world tube in $V_{4}$.

Let $\sigma _{0}$ be such a scale that $\alpha \left( \sqrt{2\sigma }\right)
\lesssim 1$ at $\left\vert \sigma \right\vert <\sigma _{0}$ and $\alpha
\left( \sqrt{2\sigma }\right) \ll 1$ at $\sigma \gg \sigma _{0}$. This
supposition agrees with Eq. (\ref{a2.4}). Then at the large mass $\mu \gg
\sqrt{2\sigma _{0}}$ the world tube of the particle distinguishes from a
straight line slightly. At $\mu =0$ the world tube coincides with the light
cone that associates usually with world lines of massless photons.

In the linear model the mass $\mu $ of a free particle is not a geometric
characteristic, because it cannot be determined by the shape of the free
particle world line. But in the tubular model the mass $\mu $ is a geometric
characteristic, because the free particle world tube determines the momentum
completely including the mass $\mu $. In this case the mass is measured in
units of length, and some universal constant $b$ defined by the relation%
\begin{equation}
m=b\mu ,\qquad \left[ b\right] =\mathrm{g/cm}  \label{a2.8}
\end{equation}%
is necessary. Here, $m$ is the mass measured in grams, and $\mu $ is the
mass measured in centimeters. In the linear model any two infinitesimally
close points $P_{0}^{\prime },P_{1}^{\prime }$ ; on $\mathcal{T}_{P_{0}P_{1}}
$, determine the momentum $m\mathbf{P}_{0}^{\prime }\mathbf{P}_{1}^{\prime
}/\left\vert \mathbf{P}_{0}^{\prime }\mathbf{P}_{1}^{\prime }\right\vert \ $
(at given mass $m$). For instance, if $P_{0}^{\prime },P_{1}^{\prime }\in
\mathcal{T}_{P_{0}P_{1}}$, then these points determine the world line $%
\mathcal{T}_{P_{0}^{\prime }P_{1}^{\prime }}=\mathcal{T}_{P_{0}P_{1}}$. The
momentum $m\mathbf{P}_{0}^{\prime }\mathbf{P}_{1}^{\prime }/\left\vert
\mathbf{P}_{0}^{\prime }\mathbf{P}_{1}^{\prime }\right\vert $ can be
ascribed to the point $P_{0}^{\prime }$, because $P_{1}^{\prime }\rightarrow
P_{0}^{\prime }$.

In the tubular model two points $P_{0}^{\prime },P_{1}^{\prime }\in \mathcal{%
T}_{P_{0}P_{1}}$ $\left( S(P_{0}^{\prime },P_{1}^{\prime
})=S(P_{0},P_{1})=\mu \right) $ determine, generally, another world tube $%
\mathcal{T}_{P_{0}^{\prime }P_{1}^{\prime }}$ which does not coincide with $%
\mathcal{T}_{P_{0}P_{1}}$. In other words, a measurement of momentum on the
tube $\mathcal{T}_{P_{0}P_{1}}$ changes the world tube (converts $\mathcal{T}%
_{P_{0}P_{1}}$ into $\mathcal{T}_{P_{0}^{\prime }P_{1}^{\prime }}$). This
fact associates with the measurement in quantum mechanics, where the
measurement changes the state of the system. Besides, the momentum $\mathbf{P%
}_{0}\mathbf{P}_{1}$ can be ascribed only to the tube segment $\mathcal{T}%
_{[P_{0}P_{1}]}$, but not to the point $P_{0}$. This fact associates with
the quantum mechanics, where the particle momentum cannot be measured
instantaneously.

Now let us try to use another approach. Let the world tube of a particle of
the mass $\mu $ be described as a broken tube%
\begin{equation}
\mathcal{T}_{\mathrm{br}}\left( \mu \right) =\dbigcup\limits_{i}\mathcal{T}%
_{[P_{i}P_{i+1]}},\qquad S\left( P_{i},P_{i+1}\right) =\mu ,\qquad i=0,\pm
1,\pm 2,...  \label{a2.9}
\end{equation}%
The segments $\mathcal{T}_{[P_{i}P_{i+1]}}$ are links of the broken tube. In
order for the particle to be free, the broken tube $\mathcal{T}_{\mathrm{br}%
}\left( \mu \right) $ should satisfy some conditions. In the case of linear
model and infinitesimally short links these conditions reduce to the first
Newton law%
\begin{equation}
\frac{dp_{i}}{d\tau }=0,\qquad i=0,1,2,3  \label{a2.10}
\end{equation}%
where $p_{i}$ is the particle momentum in Galilean frame, $\tau $ is some
parameter along world line.

In the case of finite links Eq. (\ref{a2.10}) can be rewritten in the form%
\begin{equation}
\mathbf{P}_{i}\mathbf{P}_{i-1}\downarrow \uparrow \mathbf{P}_{i}\mathbf{P}%
_{i+1},\qquad i=0,\pm 1,\pm 2,...  \label{a2.11}
\end{equation}%
Equation (\ref{a2.11}) can be used in the case when the links of the broken
world line are not infinetesimally short and satisfy Eq. (\ref{a2.9}). By
means of Eqs. (\ref{a1.3}), (\ref{a1.4}), (\ref{a1.6}), (\ref{a2.9}), the
relations (\ref{a2.11}) are reduced to the form%
\begin{equation}
S\left( P_{i-1},P_{i+1}\right) =2\mu ,\qquad i=0,\pm 1,\pm 2,...
\label{a2.12}
\end{equation}

In the linear model Eqs. (\ref{a2.9})) (\ref{a2.12}) are equivalent to Eq.(%
\ref{a2.10}) and any link $\mathcal{T}_{[P_{i}P_{i+1}]}$ determines
unambiguously the whole broken tube $\mathcal{T}_{\mathrm{br}}\left( \mu
\right) $ which is a straight line. Thus Eqs. (\ref{a2.9}), (\ref{a2.12})
can be considered as a wording of the first Newton law which is supposed to
be valid for any tubular model of the space-time also.

In the general case the fixed momentum $\mathbf{P}_{0}\mathbf{P}_{1}$, and
Eqs.(\ref{a2.9}), (\ref{a2.12}) do not determine unambiguously the broken
tube $\mathcal{T}_{\mathrm{br}}\left( \mu \right) $. It means that the
broken tube $\mathcal{T}_{\mathrm{br}}\left( \mu \right) $ should be
considered as a random tube. Statistical methods should be used for its
description.

\section{Statistical description}

The statistical description we shall use is the conventional description in
terms of the statistical ensemble. But it contains some nonconventional
details. Statistical descriptions of classical stochastic systems and
quanta1 ones distinguish. The first is based on the concept of probability
and the second is based on the concept of the probability amplitude. But
statistical descriptions of both types of dynamical systems have common
features that can be presented without using concepts of probability or
probability amplitude.

Such a description is based on the concept of the statistical ensemble as a
dynamical system.

\textit{Definition 3.1}: A dynamical system $\mathcal{S}$ whose state $X$
evolves according to some dynamical equations is a deterministic dynamical
system.

\textit{Definition 3.2}: A dynamical system $\mathcal{S}$ is a
nondeterministic dynamical system (or stochastic system), if there exist no
dynamical equations that determine its state evolution.

A characteristic property of stochastic system is an irreproducibility of
measurements. It means that repeating measurements of the same quantity $%
\mathcal{R}$ in the same state $X$ of the stochastic system $\mathcal{S}$
give different values $R_{1}$, $R_{2}$,... . Practically one cannot study
stochastic systems without reducing them to deterministic ones. The way of
such reduction is determined by the statistical principle.

\textit{Statistical principle}: A set $\mathcal{S}$ of $N$ $(N\rightarrow
\infty )$ like independent dynamical systems $\mathcal{S}$ (stochastic or
not) is a deterministic dynamical system. This system is called the
statistical ensemble. The systems $\mathcal{S}$ are called the ensemble
elements.

A result of the measurement of the quantity $\mathcal{R}$ in the ensemble $%
\mathcal{E}$ is a distribution $f(R)$ which is reproducible at other
measurements, even though the measurement of $\mathcal{R}$ in a single
ensemble system $\mathcal{S}$ is irreproducible.

All attributes of the dynamical system: energy, momentum, angular momentum,
their densities, Lagrangian, and other dynamical quantities can be ascribed
to the ensemble $\mathcal{E}$.

The dynamical equations of the ensemble $\mathcal{E}$ are insensitive to the
number $N$ of the ensemble elements. Using independence of dynamical
equations on $N$, formally one can set $N=1$ and use the ensemble consisting
of one element. Such a procedure will be referred to as an ensemble
projection onto one system. Any additive quantity $\mathcal{R}$ (energy,
momentum, etc.) of the ensemble $\mathcal{E}$ \textquotedblleft consisting
of one element\textquotedblright\ can be considered as the mean value of
this quantity $\mathcal{R}$ for the stochastic system $\mathcal{S}$.

As far as any statistical ensemble $\mathcal{E}^{\prime }$ is a dynamical
system, it can be an element of other ensemble $\mathcal{E}$. Let us
consider a Hamiltonian system $\mathcal{S}$ whose state is described by
canonical variables $(\mathbf{x,p})$ and a hierarchy of ensembles that can
be constructed on the base of this system $\mathcal{S}$. The state of the
general ensemble $\mathcal{E}$ consisting of systems $\mathcal{S}$ is
described by a non-negative function $F(\mathbf{x,p})$. The distribution
function $F$ evolves according to the Liouville equation%
\begin{equation}
\mathcal{E}:\qquad \frac{\partial F}{\partial t}+\frac{\partial H}{\partial
\mathbf{p}}\frac{\partial F}{\partial \mathbf{x}}-\frac{\partial H}{\partial
\mathbf{x}}\frac{\partial F}{\partial \mathbf{p}}=0  \label{a3.1}
\end{equation}%
where $H=H(\mathbf{x,p})$ is the Hamiltonian function of $\mathcal{S}$. The
special (pure) statistical ensemble $\mathcal{E}_{\rho ,S}$ whose state is
described by functions $\rho ,S$ given in the coordinate space is an
important special case of the statistical ensemble $\mathcal{E}$. In this
case

\begin{equation}
F_{\rho ,S}\left( \mathbf{x,p}\right) =\rho \left( \mathbf{x}\right) \delta
\left( \mathbf{p-\nabla }S\left( \mathbf{x}\right) \right)  \label{a3.2}
\end{equation}

Substitution of Eq. (\ref{a3.2}) into Eq. (\ref{a3.1}) leads to equations
describing the state $\rho ,S$ evolution of the pure ensemble $\mathcal{E}%
_{\rho ,S}$%
\begin{equation}
\mathcal{E}_{\rho ,S}:\left\{
\begin{array}{c}
\frac{\partial \rho }{\partial t}+\mathbf{\nabla }\left( \rho \frac{\partial
H}{\partial \mathbf{p}}\left( \mathbf{x\nabla }S\right) \right) =0 \\
\frac{\partial S}{\partial t}+H\left( \mathbf{x\nabla }S\right) =0%
\end{array}%
\right.  \label{a3.3}
\end{equation}%
Equations (\ref{a3.1}) and (\ref{a3.3}) have the system of Hamiltonian
equations%
\begin{equation}
\mathcal{S}:\qquad \frac{d\mathbf{x}}{dt}=\frac{\partial H}{\partial \mathbf{%
p}},\qquad \frac{d\mathbf{p}}{dt}=-\frac{\partial H}{\partial \mathbf{x}}
\label{a3.4}
\end{equation}%
as characteristics.

The ensemble $\mathcal{E}_{\rho ,S}$ described by the system (\ref{a3.3})
will be referred to as a pure ensemble, and the ensemble $\mathcal{E}$
described by Eq. (\ref{a3.1}) will be referred to as mixed one. These terms
are taken from the quantum mechanics, where the pure ensemble is described
by the wave function $\psi $. The mixed ensemble is such an ensemble whose
elements are pure ensembles. (The mixed ensemble consists of pure
ensembles.) In the given case the ensemble $\mathcal{E}_{\rho ,S}$ can be
described by means of the wave function%
\begin{equation}
\psi =\sqrt{\rho }\exp \left( iS/\hbar \right)   \label{a3.5}
\end{equation}%
where $\hbar $ is the Planck's constant. The dynamical equation for $\psi $
can be obtained from Eqs. (\ref{a3.3}), (\ref{a3.5}). It is not linear,
generally speaking. Here, the Planck's constant is used as a universal
constant having dimensionality of action. Its use in the classical
statistical ensemble description does not produce any quantum effects. The
ensemble $\mathcal{E}$ described by Eq. (\ref{a3.1}) can be treated as a
mixed one, because it can be considered as consisting of elements $\mathcal{E%
}_{\rho ,S}$, i.e., of pure ensemble describing by the wave function (\ref%
{a3.5}). Thus the dynamical systems $\mathcal{S}$, $\mathcal{E}_{\rho ,S}$, $%
\mathcal{E}$ form a hierarchy: $\mathcal{S}$ is an element of the ensemble $%
\mathcal{E}_{\rho ,S}$, $\mathcal{E}_{\rho ,S}$ is an element of the
ensemble $\mathcal{E}$.

Before considering the statistical description of broken world tubes $%
\mathcal{T}_{\mathrm{br}}$ let us investigate a pure statistical ensemble $%
\mathcal{E}_{\rho ,S}$ of relativistic classical particles in $E_{4}$.

The ensemble $\mathcal{E}_{\rho ,S}$ is a Hamiltonian system whose state is
described by canonically conjugate variables $\rho (\mathbf{x}),S(\mathbf{x}%
) $ numerated by spatial coordinates $\mathbf{x}$. Its Hamiltonian functional%
\begin{equation}
\mathcal{H}\left( \rho ,S\right) =\int \rho c\sqrt{m^{2}c^{2}+\left( \mathbf{%
\nabla }S\right) ^{2}}d\mathbf{x}  \label{a3.6}
\end{equation}%
generates dynamical equations%
\begin{equation}
\frac{\partial \rho \left( \mathbf{x}\right) }{\partial t}=\frac{\delta
\mathcal{H}}{\delta S\left( \mathbf{x}\right) },\qquad \frac{\partial
S\left( \mathbf{x}\right) }{\partial t}=-\frac{\delta \mathcal{H}}{\delta
\rho \left( \mathbf{x}\right) },  \label{a3.7}
\end{equation}%
where $\delta /\delta \rho ,\delta /\delta S$ are variational derivatives.

Let us note the following mathematical fact. Replacing $m^{2}$ by%
\begin{equation}
m_{q}^{2}=m^{2}+\left( \frac{\hbar }{2c}\mathbf{\nabla }\log \rho \right)
^{2},\qquad m^{2}=\mathrm{const,}\qquad \left( \frac{\hbar }{2c}\mathbf{%
\nabla }\log \rho \right) ^{2}\ll m^{2}  \label{a3.8}
\end{equation}%
in Eq. (\ref{a3.6})) one obtains dynamical equations that are equivalent in
the nonrelativistical approximation to the Schrodinger equation for a free
particle%
\begin{equation}
i\hbar \frac{\partial \psi }{\partial t}=-\frac{\hbar ^{2}}{2m}\mathbf{%
\nabla }^{2}\psi  \label{a3.9}
\end{equation}%
with the wave function%
\begin{equation}
\psi =\sqrt{\rho }\exp \left( \frac{i}{\hbar }\left( mc^{2}t+S\right) \right)
\label{a3.10}
\end{equation}%
Here, the first term $mc^{2}t$ in exponent compensates the rest mass term of
the relativistical variable $S$.

Finally, let us produce transformation to \textquotedblleft geometrical
variables\textquotedblright
\begin{equation}
S=bc\tilde{S},\quad m=b\mu ,\quad t=\tau /c,\quad \hbar =bc\sigma _{0},\quad
\sigma _{0}=\mathrm{const}  \label{a3.11}
\end{equation}%
where $b$ is defined by Eq. (\ref{a2.8}). Then the Hamiltonian of the
quantum particles ensemble takes the form%
\begin{equation}
\mathcal{H}\left( \rho ,\tilde{S}\right) =\int \sqrt{\mu _{q}^{2}+\left(
\mathbf{\nabla }\tilde{S}\right) ^{2}}\rho d\mathbf{x}  \label{a3.12}
\end{equation}%
\begin{equation}
\mu _{q}^{2}=gm^{2}+\left( \frac{\sigma _{0}}{2}\mathbf{\nabla }\log \rho
\right) ^{2}  \label{a3.13}
\end{equation}%
One can see from Eq. (\ref{a3.12}) that quantum effects arise on account of
dependence of the mass $\mu $ on the ensemble state $(\rho ,S)$.

Unlike the Hamiltonian (\ref{a3.6}) the Hamiltonian (\ref{a3.12}) cannot be
considered as a sum of Hamiltonians of independent systems, because the mass
$\mu _{q}$ depends on the ensemble state $\rho ,S$. It means that in the
hierarchy $\mathcal{S}$, $\mathcal{E}_{\rho ,S}$, $\mathcal{E}$ the state of
a single system $\mathcal{S}$ does not satisfy any dynamical equation, i.e.,
$\mathcal{S}$ is a stochastic system. The dynamical system $\mathcal{E}%
_{\rho ,S}$ is the first deterministic dynamical system in the hierarchy $%
\mathcal{S}$, $\mathcal{E}_{\rho ,S}$, $\mathcal{E}$. Investigation of the
stochastic system $\mathcal{S}$ is meaningless. The first dynamical system
on the hierarchy that has to be investigated is the pure ensemble $\mathcal{E%
}_{\rho ,S}$.

\section{Oriented mass}

\textit{Definition 4.2}: The oriented mass $\mu _{o}(P)$ of the tube $%
\mathcal{T}_{P_{0}P_{1}}$ in $V_{4}$ is defined by the equation

\begin{equation}
\mu _{o}=\frac{\left( \mathbf{P}_{0}\mathbf{P}_{1}\cdot \mathbf{P}_{0}%
\mathbf{P}\right) }{\left\vert \mathbf{P}_{0}\mathbf{P}_{1}\right\vert }%
=\left\{
\begin{array}{cc}
\mu & P\in \mathcal{T}_{[P_{0}P_{1}} \\
-\mu & P\in \mathcal{T}_{P_{0}]P_{1}}%
\end{array}%
\right. ,\quad \mu =S\left( P_{0},P_{1}\right)  \label{a4.1}
\end{equation}%
In the case of the linear model $\mu _{q}(P)$ degenerates into%
\begin{equation}
\mu _{o}=p_{i}u^{i}=\pm \mu  \label{a4.2}
\end{equation}%
where $p_{i}=\mathbf{P}_{0}\mathbf{P}_{1}$ and $u^{i}=\mathbf{P}_{0}\mathbf{P%
}/\left\vert \mathbf{P}_{0}\mathbf{P}\right\vert $ are correspondently
four-momentum and four-velocity of the particle. The sign of the mass
describes the mutual orientation of four-momentum and four-velocity. It is
different for particle and antiparticle.

\textit{Definition 4.2}: The associated oriented mass $\mu _{oE}$ of $%
\mathcal{T}_{P_{0}P_{1}}$ in $V_{4}$ is the oriented mass of $\mathcal{T}%
_{P_{0}P_{1}}$ considered from the standpoint of the associated $\sigma $
space $V_{4}$ :%
\begin{equation}
\mu _{oE}=\frac{\left( \mathbf{P}_{0}\mathbf{P}_{1}\cdot \mathbf{P}_{0}%
\mathbf{P}\right) _{\mathrm{E}}}{\left\vert \mathbf{P}_{0}\mathbf{P}%
_{1}\right\vert _{\mathrm{E}}},\qquad P\in \mathcal{T}_{P_{0}P_{1}},
\label{a4.3}
\end{equation}%
where index \textquotedblleft E\textquotedblright\ means that the scalar
product is calculated in the associated space $E_{4}$. Vectors $\mathbf{P}%
_{0}\mathbf{P}_{1}$ and $\mathbf{P}_{0}\mathbf{P}$, $P\in \mathcal{T}%
_{P_{0}P_{1}}$ are not collinear in $E_{4}.$ Vector of four-momentum and
that of four-velocity are not collinear in $E_{4}$.

In the coordinate system, where spatial components of the four-momentum
vanish,%
\begin{equation}
p_{i}=\left( \mu _{\mathrm{E}},\mathbf{0}\right) .\qquad u^{i}=\left( 1/%
\sqrt{1-\mathbf{\beta }^{2}},\mathbf{\beta /}\sqrt{1-\mathbf{\beta }^{2}}%
\right)  \label{a4.4}
\end{equation}%
one has%
\begin{equation}
\mu _{o\mathrm{E}}=p_{i}u^{i}=\mu _{\mathrm{E}}/\sqrt{1-\mathbf{\beta }^{2}}%
\simeq \mu _{\mathrm{E}}\left( 1-\mathbf{\beta }^{2}/2\right) ,\qquad
\mathbf{\beta }^{2}\ll 1  \label{a4.5}
\end{equation}%
Here $\mathbf{\beta }$ is a velocity in the coordinate system, where $%
\mathbf{p}=0$, and $\theta $ = arctanh$\beta $ is the angle between vectors $%
p_{i}$ and $u^{i}$

For the broken world tube (\ref{a2.9}) of the mass $\mu $ the oriented
associated mass is defined as follows%
\begin{equation}
\mu _{o\mathrm{E}}\left( P\right) =\frac{\left( \mathbf{P}_{i}\mathbf{P}%
_{i+1}\cdot \mathbf{P}_{i}\mathbf{P}\right) _{\mathrm{E}}}{\left\vert
\mathbf{PP}_{1}\right\vert _{\mathrm{E}}},\qquad P\in \mathcal{T}%
_{P_{i}P_{i+1}}\subset \mathcal{T}_{\mathrm{br}}\left( \mu \right)
\label{a4.6}
\end{equation}

\section{Determination of mapping function $f$}

Let us consider a nonrelativistic ensemble $\mathcal{E}$ of broken tubes $%
\mathcal{T}_{\mathrm{br}}\left( \mu \right) $ in the tubular model $V_{4}$,
. Let this ensemble be described in the associated space $E_{4}$. It is
supposed that among the ensembles $\mathcal{E}$ there are pure
nonrelativistic ensembles $\mathcal{E}_{\rho ,S}$ having the following
properties.

(1) The state of the pure ensemble $\mathcal{E}_{\rho ,S}$ is described in $%
E_{4}$ by the particle concentration $\rho $ and momentum $p_{i}=\left\{
\sqrt{\mu _{\mathrm{E}}^{2}+\left( \mathbf{\nabla }S\right) ^{2}},\mathbf{%
\nabla }S\right\} $, $(p_{i}p^{i}=\mu _{\mathrm{E}}^{2})$.

(2) The ensemble $\mathcal{E}_{\rho ,S}$ state evolution is described by the
Hamiltonian functional (\ref{a3.12})) where $\mu _{q}$ coincides with the
oriented associated mass $\mu _{o\mathrm{E}}$ which is some function of the
ensemble state $(\rho ,S)$. $\mu _{o\mathrm{E}}$ is defined by Eqs. (\ref%
{a4.4})) (\ref{a4.5}). $\mathbf{\beta }$ is the mean velocity at the point $P
$ in the coordinate system, where the mean momentum $\mathbf{p}$ vanishes.

The $\mathbf{\beta }$ is some function of the ensembIe state $\left( \rho
,S\right) $. It must be calculated basing on the tubular space-time model $%
V_{4}$. That mode1 $V_{4}$, where

\begin{equation}
\mathbf{\beta }=-\frac{\sigma _{0}}{2\mu _{\mathrm{E}}}\frac{\mathbf{\nabla }%
\rho }{\rho }  \label{a5.1}
\end{equation}%
and, hence,$\mu _{o\mathrm{E}}$ coincides with $\mu _{q}^{2}$ in Eq. (\ref%
{a3.13}) is an optima1 one, because the ensemble $\mathcal{E}_{\rho ,S}$ of
broken tubes coincides with quanta1 ensemble (\ref{a3.12}), and the tubes $%
\mathcal{T}_{\mathrm{br}}\left( \mu \right) $ of this $V_{4}$, can be
treated as world tubes of real particles. Let us note that Eq. (\ref{a5.1})
coincides with the expression for the mean velocity of the ensemble of
Brownian particles, where particle concentration is $\rho $ and diffusion
coefficient is $D=0.5\sigma _{0}/\mu _{\mathrm{E}}$.

Let us calculate the mean velocity $\mathbf{\beta }$ in the ensemble $%
\mathcal{E}_{\rho ,S}$. Let us attribute the momentum $\mathbf{p}$, defined
by the vector $\mathbf{P}_{0}\mathbf{P}_{1}$, to the point $P$ which is
placed in the middle of the segment $[P_{0},P_{1}]$ in $E_{4}$ (see Fig. 2):%
\begin{equation}
S_{\mathrm{E}}\left( P_{0},P\right) =S_{\mathrm{E}}\left( P,P_{1}\right) =%
\frac{1}{2}S_{\mathrm{E}}\left( P_{0},P_{1}\right) =\frac{1}{2}g\left( \mu
\right) =\frac{1}{2}\mu _{\mathrm{E}}  \label{a5.2}
\end{equation}

It is supposed that the tube segment $\mathcal{T}_{\left[ P_{0}^{\prime
}P_{1}^{\prime }\right] }$ of any broken tube which contains the point $P$, $%
P\in \mathcal{T}_{\left[ P_{0}^{\prime }P_{1}^{\prime }\right] }$
contributes into the value $u^{i}(P)$ of the four-velocity at $P$. This
contribution is proportional to $x^{i}(P)-x^{i}(P^{\prime })$, where $%
x^{i}(Q)$ are Galilean coordinates of the point $Q$ in $E_{4}$. This
coordinate system is supposed to be chosen in such a way, that%
\begin{equation}
x^{i}\left( P\right) =0,\qquad p^{i}=x^{i}(P_{1})-x^{i}(P_{0})=\left( \mu _{%
\mathrm{E}},\mathbf{0}\right)   \label{a5.3}
\end{equation}%
For simplicity, contributions of only those segments $\mathcal{T}_{\left[
P_{0}^{\prime }P_{1}^{\prime }\right] }$ are taken into account for which%
\begin{equation}
S\left( P,P_{0}^{\prime }\right) =S\left( P,P_{1}^{\prime }\right) ,\qquad
P\in \mathcal{T}_{\left[ P_{0}^{\prime }P_{1}^{\prime }\right] }
\label{a5.4}
\end{equation}%
It means that in $E_{4}$ the point $P\in \mathcal{T}_{\left[
P_{0}^{\prime }P_{1}^{\prime }\right] }$ is maximally remoted from
the segment $\left[ P_{0}^{\prime }P_{1}^{\prime }\right] $. Let
$P^{\prime }$ be the middle of
the segment $\left[ P_{0}^{\prime }P_{1}^{\prime }\right] $ in $E_{4}$, i.e.,%
\begin{equation}
S_{\mathrm{E}}\left( P_{1}^{\prime },P^{\prime }\right) =S_{\mathrm{E}%
}\left( P^{\prime },P_{0}^{\prime }\right) =\frac{1}{2}S_{\mathrm{E}}\left(
P_{0}^{\prime },P_{1}^{\prime }\right) =\frac{1}{2}g\left( \mu \right) =%
\frac{1}{2}\mu _{\mathrm{E}}  \label{a5.5}
\end{equation}%
then%
\begin{equation}
S_{\mathrm{E}}\left( P,P^{\prime }\right) =iR_{\mathrm{E}}\sqrt{\frac{1}{4}%
g^{2}\left( \mu \right) -g^{2}\left( \mu /2\right) }  \label{a5.6}
\end{equation}%

\newpage
\begin{figure}[ptb]
\begin{center}
\includegraphics [height=7cm,keepaspectratio]{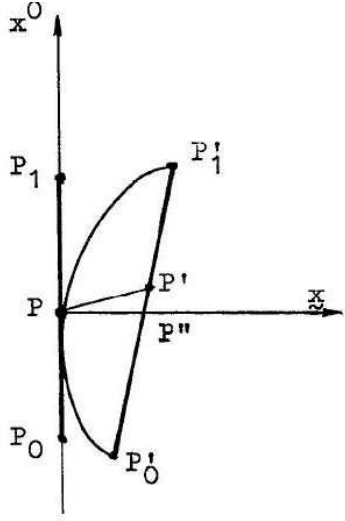}
\caption{Momentum $p_i$, defined by the vector
$\mathbf{P}_0\mathbf{P}_1$ to the point $P$ which is placed in the
middle of the segment $[P_0P_1]$ in $E_4$}\label{f2}
\end{center}
\end{figure}

The flux density $j^{i}(P^{\prime \prime })$ is defined by the relation%
\begin{equation}
j_{k}(P^{\prime \prime })=\left( \rho \left( P^{\prime \prime }\right) ,%
\frac{\rho \left( P^{\prime \prime }\right) \mathbf{\nabla }S\left(
P^{\prime \prime }\right) }{\sqrt{\mu _{\mathrm{E}}^{2}+\left( \mathbf{%
\nabla }S\left( P^{\prime \prime }\right) \right) ^{2}}}\right)
\label{a5.7}
\end{equation}%
and vector $j^{i}(P^{\prime })$ is parallel to $p^{i}(P^{\prime })$ in $E_{4}
$.

Let%
\begin{equation}
y^{i}=x^{i}\left( P^{\prime \prime }\right) ,\qquad i=0,1,2,3;\qquad y^{0}=0
\label{a5.8}
\end{equation}%
The flux density $j^{i}(y)$ is supposed to change on the characteristic
length $L=\mu /\varepsilon $, $\varepsilon \ll 1$. Then%
\begin{equation}
j^{i}\left( P^{\prime \prime }\right) =j^{i}\left( P\right)
+j_{,k}^{i}\left( P\right) y^{k}+O\left( \varepsilon ^{2}\right) ,\quad
j_{,k}^{i}\equiv \partial _{k}j^{i},\quad \mu \partial
_{k}j^{i}/j^{0}=O\left( \varepsilon \right)   \label{a5.9}
\end{equation}%
One can calculate that%
\[
x^{i}\left( P^{\prime }\right) =y^{i}+q^{i}\left( P^{\prime }\right) l_{%
\mathrm{E}},\quad l_{\mathrm{E}}=S_{\mathrm{E}}\left( P^{\prime },P"\right) =%
\sqrt{R_{\mathrm{E}}^{2}-\mathbf{y}^{2}}
\]%
\begin{equation}
q^{i}\left( P^{\prime }\right) \equiv \frac{j^{i}\left( P^{\prime }\right) }{%
\left\vert \mathbf{j}\left( P^{\prime }\right) \right\vert }=\frac{%
j^{i}\left( P\right) +j_{,k}^{i}\left( P\right) \left( y^{k}+\delta
_{0}^{k}l_{\mathrm{E}}\right) }{j^{0}\left( P\right) }+O\left( \varepsilon
^{2}\right)   \label{a5.10}
\end{equation}%
Calculation shows that Eqs. (\ref{a5.4}), (\ref{a5.5}) are equivalent to%
\begin{equation}
R_{\mathrm{E}}^{2}-\mathbf{y}^{2}+O\left( \varepsilon ^{2}\right) =0
\label{a5.11}
\end{equation}

Coordinates $x^{i}(P_{0}^{\prime })$ are determined by the relation%
\begin{eqnarray}
x^{i}(P_{0}^{\prime }) &=&y^{i}-q^{i}\left( P^{\prime }\right) \left( S_{%
\mathrm{E}}\left( P_{0}^{\prime },P^{\prime }\right) -l_{\mathrm{E}}\right)
\nonumber \\
&=&y^{i}-\frac{j^{i}+j_{,\alpha }^{i}y^{\alpha }+j_{,0}^{i}l_{\mathrm{E}}}{%
j^{0}}\left( \frac{g\left( \mu \right) }{2}-l_{\mathrm{E}}\right)
\label{a5.12}
\end{eqnarray}

Let us define the mean four-velocity by means of the expression%
\begin{equation}
u^{i}\left( P\right) =A\int \left( x^{i}\left( P\right) -x^{i}\left(
P_{0}^{\prime }\right) \right) j^{i}\left( P^{\prime \prime }\right) \delta
\left( R_{\mathrm{E}}^{2}-\mathbf{y}^{2}\right) d\mathbf{y}+O\left(
\varepsilon ^{e}\right)   \label{a5.13}
\end{equation}%
where the normalization factor $A$ is defined by the relation%
\begin{equation}
u^{i}\left( P\right) u_{i}\left( P\right) =1  \label{a5.14}
\end{equation}

One obtains%
\begin{eqnarray}
u^{0} &=&A\int \frac{1}{2}g\left( \mu \right) \left( 1+\frac{j_{,0}^{0}l_{%
\mathrm{E}}}{j^{0}}+\frac{j_{,\alpha }^{0}y^{\alpha }}{j^{0}}\right) \left(
j^{0}+j_{,\alpha }^{0}y^{\alpha }\right) \delta \left( R_{\mathrm{E}}^{2}-%
\mathbf{y}^{2}\right) d\mathbf{y}  \nonumber \\
&=&\pi R_{\mathrm{E}}^{2}Ag\left( \mu \right) j^{0}+O\left( \varepsilon
^{2}\right)   \label{a5.15}
\end{eqnarray}%
\begin{eqnarray}
u^{\alpha } &=&A\int \left( -y^{\alpha }+\frac{1}{2}g\left( \mu \right)
\right) \left( \frac{j_{,0}^{\alpha }l_{\mathrm{E}}}{j^{0}}+\frac{j_{,\alpha
}^{\alpha }y^{\alpha }}{j^{0}}\right) \left( j^{0}+j_{,\alpha }^{0}y^{\alpha
}\right) \delta \left( R_{\mathrm{E}}^{2}-\mathbf{y}^{2}\right) d\mathbf{y}
\nonumber \\
&=&-\frac{2\pi }{3}AR_{\mathrm{E}}^{3}j_{,\alpha }^{0}+O\left( \varepsilon
^{2}\right)   \label{a5.16}
\end{eqnarray}%
It follows from Eqs. (\ref{a5.15}), (\ref{a5.16})%
\begin{equation}
\mathbf{\beta }=\frac{\mathbf{u}}{u^{0}}=-\frac{2}{3}\frac{R_{\mathrm{E}}^{2}%
}{\mu _{\mathrm{E}}}\frac{\mathbf{\nabla }\rho }{\rho }  \label{a5.17}
\end{equation}%
Comparing Eqs. (\ref{a5.1}) and (\ref{a5.17}), one obtains by means of Eq. (%
\ref{a2.7})%
\begin{equation}
R_{\mathrm{E}}^{2}=\frac{1}{2}g^{2}\left( \mu \right) -g^{2}\left( \frac{\mu
}{2}\right) =\frac{3}{4}\sigma _{0}  \label{a5.18}
\end{equation}%
Let Eq. (\ref{a5.18}) take place for any rnass $\mu >\mu _{e}$, where $\mu
_{e}$ is the electron mass, i.e., the mass of the lightest massive
particles. If the particle mass $\mu =0$, then the nonrelativistic
approximation which was used for obtaining Eq. (\ref{a5.18}) does not exist.
One has a solution of the functional equation (5.18):%
\begin{equation}
\mu _{\mathrm{E}}^{2}=g^{2}\left( \mu \right) =\mu ^{2}-\sigma _{0},\qquad
\mu \geq \mu _{e}  \label{a5.19}
\end{equation}%
It follows from Eqs. (\ref{a2.3}), (\ref{a5.19})%
\begin{equation}
\sigma =\sigma _{\mathrm{E}}+\frac{\sigma _{0}}{2},\qquad \sigma _{0}=\frac{%
\hbar }{bc},\qquad \sigma \geq \frac{\mu _{e}^{2}}{2}  \label{a5.20}
\end{equation}

On the other hand, according to Eq. (\ref{a1.1}), $\sigma $ and $\sigma _{%
\mathrm{E}}$ vanish simultaneously. Substituting $\mu =\mu _{e}$ into Eq. (%
\ref{a5.19}) and using the relation $\mu _{\mathrm{E}}^{2}=g^{2}\left( \mu
_{e}\right) >0$, one obtains
\begin{equation}
\mu _{e}>\sigma _{0}  \label{a5.21}
\end{equation}%
It follows from Eqs. (\ref{a5.21}), (\ref{a3.11})%
\begin{equation}
b<10^{-16.5}\mathrm{g/cm,\qquad }\sqrt{\sigma _{0}}<10^{-10.5}\mathrm{cm}
\label{a5.22}
\end{equation}%
The maximal section radius of the broken tube%
\begin{equation}
R_{\mathrm{E}}-\frac{\sqrt{3\sigma _{0}}}{2}<10^{-10.4}\mathrm{cm}
\label{a5.23}
\end{equation}%
is the same for all massive particles. The quantity $\sqrt{\sigma _{0}}$ can
be treated as an elementary length. The obtained relation (\ref{a5.20})
determines the optimal tubular model of the space-time. This space-time
model is responsible for quantum effects.

Let us note the relation (\ref{a5.20}) determines the mapping function $f$
only for the values of argument $\sigma _{\mathrm{E}}\geq \left( \mu
_{e}^{2}-\sigma _{0}\right) /2>0.$For $\sigma _{\mathrm{E}}<\left( \mu
_{e}^{2}-\sigma _{0}\right) /2$ the mapping function $f$ remains
indeterminate.

\section{Discussion}

The tubular model (\ref{a5.20}) is a more attractive model of the space-time
than the linear one, because it explains some more quantum properties. In
particular, it permits random world tubes to exist, the pure ensemble of
these tubes being described by the Schr\"{o}dinger equation.

Of course, the quantum mechanics principIes are not exhausted by the Schr%
\"{o}dinger equation. Besides the linearity principle they include more the
rule of computation of the physical quantities average values. Using the
statistical principle \cite{R73,R80} this rule can be obtained. According to
the statistical principle the statistical ensemble is a dynamical system
consisting of many dynamical systems or stochastical ones. At a proper
normalization all additive physical quantities (energy, momentum, angular
momentum) of the ensemble can be considered as the mean values of the
corresponding quantities for the ensemble elements. Besides, for the
nonrelativistical ensemble the $\rho $ can be treated as a probability
density of the particle position. It allows us to calculate the mean values
of the type $F(\mathbf{x})$. As regards to mean values of type $F(\mathbf{p})
$, $F(\mathbf{x,p})$ one cannot calculate them for the pure ensemble (\ref%
{a3.6})-(\ref{a3.8}). In the conventional quantum mechanics the value $F(%
\mathbf{p})$ can be calculated, but the truth of this calculated value $F(%
\mathbf{p})$ cannot be tested experimentally \cite{R77}, because the
momentum cannot be measured instantly. The system state $\psi $ changes
during the measuring time, and the measured value cannot be ascribed to any
definite state $\psi $. But the quantum mechanics always ascribes any
measured value to some definite state $\psi $.

\end{document}